\documentclass{iopart}

\usepackage[dvips]{graphicx}

\begin{document}

\title{
Strangeness and charm signatures 
%of the QGP 
in A+A collisions}

\author{Mark I. Gorenstein}

\address{Bogolyubov Institute
for Theoretical Physics,
Kiev, Ukraine}

\ead{goren@th.physik.uni-frankfurt.de}

\begin{abstract}
Two  signatures of
the quark-gluon plasma -- strangeness `enhancement'
and $J/\psi$ `suppression' --
in nucleus--nucleus collisions
are critically discussed.
A recently developed statistical coalescence model for $J/\psi$
production is presented. 
The measurements at the RHIC energies are crucial for disentangling
the different scenarios of $J/\psi$ formation.
\end{abstract}

%\newpage

\section{ Strangeness ``enhancement'' ?}

The idea of strangeness
enhancement as a quark-gluon plasma (QGP) signal
in nucleus-nucleus (A+A) collisions was formulated a long
time ago
\cite{raf:86}.
It was based on the estimate that the strangeness equilibration time
in QGP is of the same order ($\approx 10$ fm/c) as
the expected life time of the fireball formated in A+A
collisions.
Thus  in the case of QGP creation the strangeness is expected
to approach its equilibrium value in QGP.
This equilibrium value is significantly higher
than the strangeness production in
nucleon--nucleon (N+N) collisions.
Strangeness production in secondary hadronic interactions
was estimated to be negligible small (this appears to be not correct!).
Therefore, if QGP is not formed, the strangeness
yields would be expected to be much
lower than those predicted by equilibrium QGP calculations.
Thus at that time a simple and elegant signature of
QGP creation appeared:
 a transition to QGP should be signaled
by an increase of the strangeness
production to the level of
QGP equilibrium value.

How to check this prediction?
The  actual study 
%of the strangeness `enhancement'
%, due to experimental and theoretical reasons, 
has been done in the following way.
The strangeness to pion ratio,
\begin{equation}\label{strange}
E_s~=~\frac{\langle \Lambda\rangle ~+~\langle K+\overline{K}\rangle}
{\langle \pi \rangle}~,
\end{equation}
was measured and analyzed.
Different ratios of this type, $f_s=\langle K^{\pm} \rangle / \langle \pi \rangle$,
..., $\langle \Omega \rangle / \langle \pi \rangle $,
$\langle K^{\pm} \rangle / \langle N_p \rangle$, ...,$\langle \Omega \rangle
/N_p$, have been also
studied ($N_p$ is the number of nucleon participants in A+A collision).
One expected  that all these ratios should increase {\it strongly}
in A+A collisions if the QGP was formed. 
%with both the collision energy
%$\sqrt{s}$
%and the number of nucleon participants $N_p$. 
On the other hand, the strangeness to pion ratio (\ref{strange})
(as well as other specific ratios $f_s$)
increases with collision energy
$\sqrt{s}$ in N+N collision too. 
To reveal  the specific {\it strong} 
increase of the strangeness/pion ratio
in A+A collisions due to the QGP formation
the strangeness
enhancement
factor was introduced:
\begin{equation}\label{senh}
R_s~\equiv~\frac{E_s^{AA}(\sqrt{s})}{E_s^{NN}(\sqrt{s})}~,
\end{equation}
where $E_s^{AA}$ and $E_s^{NN}$ correspond respectively to
A+A and N+N collisions at the same c.m. energy $\sqrt{s}$.
Does the strangeness enhancement factor $R_s$ (\ref{senh}) become
large
if the QGP is formed? 
The confrontation of this expectation with the data
was for the first time possible in 1988 when the 
results from S and Si beams at SPS and AGS were presented.
The experiment NA35 reported that in central S+S collisions
at 200~AGeV the strangeness to pion ratio is indeed 2 times higher
than in N+N interactions at the same energy per nucleon.
But even larger enhancement ($R_s$ is about of 3)  was measured  
by E802 in Si+A collisions at AGS.
Recent data on central Au+Au collisions at low AGS energies
$4\div 10$~AGeV completed the  picture:
strangeness enhancement is observed at all energies,
and it is stronger  at lower energies, i.e. the function $R_s$
(\ref{senh}) 
{\it increases} monotonously with {\it decreasing} of $\sqrt{s}$. 
At the low AGS energies one does not expect a creation of the QGP
and therefore a substantial strangeness
enhancement is evidently of a different origin.
Thus AGS  measurements of strangeness enhancement larger
than that at SPS show clearly that the simple concept
of strangeness enhancement as a signal of QGP does not work.

Let me return to the observable $E_s^{AA}$ and its energy dependence
shown in Fig.~1.

\begin{figure}[ht]
\begin{center}
\includegraphics[height=8cm]{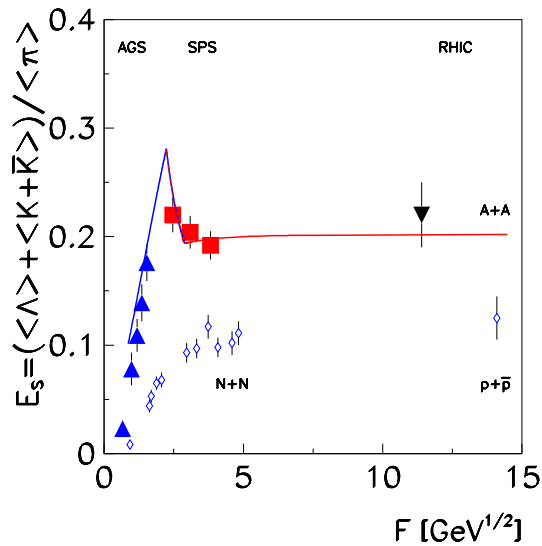}
\caption{Collision energy
($F\equiv (\sqrt{s}-2m_N)^{3/4}/\sqrt{s}^{1/4}$)
dependence of strangeness to
pion ratio (\ref{strange}) for central A+A collisions
(closed points), N+N and $p+\overline{p}$ colisions
(open points). The prediction of the statistical model
of Ref.~\cite{early} is shown by solid line.
A transition to the QGP is expected between the
AGS ($F\approx 2$~GeV$^{1/2}$) and the SPS
($F\approx 4$~GeV$^{1/2}$) energies and
leads to the non-monotonic dependence of the strangeness to
pion ratio. At high collision energies the ratio saturates at
the value characterestic for equilibrium QGP.}
\end{center}
\end{figure}

The statistical model of the early stage \cite{early} leads 
to the following predictions for strangeness production:\\
a). A non-monotonic  collision energy dependence
of the strangeness to pion ratio (\ref{strange}).
This is due to the fact
that the phase transition is expected to occur at an energy
where the strangeness to entropy (pion) ratio in the 
confined (hadron) matter is higher than in the QGP.
Therefore, a creation of the QGP in the energy region between the 
AGS and SPS would change an initial fast increase of the
ratio $E_s$ (\ref{strange}) in the hadron gas by a {\it decrease} to
the level expected in equilibrium QGP.\\
b). Very similar strangeness to pion ratio is predicted for
SPS, RHIC and LHC energies as strangeness/entropy ratio
in the QGP is almost independent of temperature (collision energy).

Preliminary SPS data in Pb+Pb at 40 and 80 AGeV and RHIC data
in Au+Au  at $\sqrt{s}=200$~GeV seem to be in agreement with the 
above conclusions (see Fig.~1).

\section{ $J/\psi$ suppression and enhancement}

A standard picture of
$J/\psi$
production in hadron and nuclear
collisions assumes a two step process: the creation of
$c\overline{c}$ pair in hard parton collisions at the very early
stage of the reaction and the subsequent formation of
a bound charmonium state. Matsui and Satz proposed \cite{Satz1}
to use
$J/\psi$ as a probe for deconfinement in the study of A+A
collisions. They argued that in QGP the colour screening dissolves
initially created $J/\psi$ mesons into $c$ and $\overline{c}$
quarks which at hadronization form open charm hadrons.
As the initial yield of $J/\psi$ is believed to have
the same A--dependence as the Drell--Yan lepton pairs,
the measurement of a weaker A--dependence of final $J/\psi$
yield ($J/\psi$ suppression)
would signal charmonium absorption and therefore
creation of QGP.
The production of charmonium states $J/\psi$ and $\psi^{\prime}$ have been
measured in A+A
collisions at CERN SPS over the last 15 years by the NA38 and NA50
Collaborations.
There are two unambiguous consequences of the standard $J/\psi$
suppression picture \cite{Satz1}.  
First, the number of
$J/\psi$ particles produced before the suppression should be directly   
proportional to the number of $c\bar{c}$ pairs, $N_{c\bar{c}}$.
Second, when the energy $\sqrt{s}$ and/or the number of nucleon
participants $N_p$
of the colliding
nuclei increase, the $J/\psi$ 
suppression becomes stronger, i.e. for the measured $J/\psi$ yield,
$\langle J/\psi \rangle$, the 
ratio
\begin{equation}\label{ratio}
R\equiv \frac{\langle J/\psi \rangle}{N_{c\overline{c}}}
\end{equation} 
decreases
with increasing of $\sqrt{s}$ and/or $N_p$.

Recently  the thermal  model \cite{Ga1}
and the statistical coalescence model \cite{Br1,Go:00}
were suggested for the charmonium production in
A+A collisions.
This statistical description makes very different assumptions about the
underlying physics which generates charmonium.  
In the statistical coalescence model  \cite{Br1,Go:00} the
$J/\psi$ particles are produced by recombination
of $c$ and $\overline{c}$ at the hadronization stage, and the
picture is much different from the standard suppression scenario.   
%Since recombination rises as the density of charm quarks increase, the
%ratio of $N_{hidden charm}/N_{open charm}$ should be a rapidly increasing
%function of energy.  
For large number of $c\overline{c}$ pairs, $N_{c\overline{c}}>>1$, the
multiplicity of $J/\psi$ due to recombination of $c$ and
$\overline{c}$
should be
roughly proportional to:
\begin{equation}\label{recom}
\langle J/\psi \rangle ~\sim ~ \frac{N^2_{c\overline{c}}}{V}~,
\end{equation}
where $V$ is the system volume.
In this case  the ratio $R$ (\ref{ratio})
is expected to increase with  increasing of $\sqrt{s}$ and/or $N_p$.
We call this $J/\psi$ {\it enhancement}.

%The two different pictures have therefore a much different dependence on
%both collision energy
%and upon the number of nucleon participants.

The equilibrium hadron gas (HG) model describes 
the hadron yields measured in A+A collisions in terms
of three parameters:  volume $V$, temperature $T$
and baryonic chemical potential $\mu_B$.
This model reproduces basic features of the data 
in the whole AGS--SPS--RHIC energy
region describing successfully 
the hadron yields
(see e.g. \cite{HG}).
For the  RHIC energies the temperature parameter $T$ is expected to be
similar to that for the SPS energies: $T=170\pm 10$~MeV. 
The baryonic chemical potential becomes small
 ($\mu_B < T$) and decreases with the collision energy.

The HG model assumes the
following formula for the hadron thermal multiplicities
in the grand canonical ensemble (g.c.e.):
\begin{equation}\label{stat}
N_j~=~\frac{d_j~V}{2\pi^2} ~
\int_0^{\infty}p^2dp~\left[\exp\left(
\frac{\sqrt{p^2+m_{j}^2} - \mu_j}{T}\right)~\pm~1\right]^{-1}~,
\end{equation}
where 
$m_j$, $d_j$ denote particle masses and
degeneracy
factors.
The particle chemical
potential $\mu_j$ in Eq.(\ref{stat})
is defined as
\begin{equation}\label{mui}
\mu_j~=~b_j\mu_B~+~s_j\mu_S~+~c_j\mu_C~,
\end{equation}
where $b_j,s_j,c_j$ denote the baryonic number strangeness and
charm of particle $j$. The baryonic chemical potential $\mu_B$
regulates the baryonic density of the HG system whereas
strange $\mu_S$ and charm $\mu_C$ chemical potentials should be found
from the requirement of zero value for the total strangeness and charm
in the system (in our consideration we neglect small effects
of a non-zero electrical chemical potential).

The total multiplicities $N_j^{tot}$ in the HG model
include the resonance decay
contributions:
\begin{equation}\label{dec}
N_j^{tot}~=~N_j ~+~\sum_R B(R \rightarrow j) N_R~,
\end{equation}
where $B(R \rightarrow j)$ are the corresponding decay
branching ratios.
The hadron yield ratios $N_j^{tot}/N_i^{tot}$
in the g.c.e. are then the functions on $T$ and $\mu_B$ variables and are
independent of the volume parameter $V$.

For the thermal multiplicities of both open charm and
charmonium states the Bose and Fermi effects are negligible,
and $m_j>>T$.
Therefore, Eq.(\ref{stat}) is simplified to:
\begin{equation}\label{gce}
N_j~
%=~ \frac{d_j
%V~e^{\mu_j/T}}{2\pi^2}~T~m^2_j~K_2\left(\frac{m_j}{T}\right)~
\cong~d_j~V~e^{\mu_j/T}~\left(\frac{m_jT}{2\pi}\right)^{3/2} \exp\left(-
\frac{m_j}{T}\right)~.
\end{equation}
%where $K_2$ is the modified Bessel function.
The HG model
gives  the $J/\psi$ yield:
\begin{equation}\label{psitot}
N_{J/\psi}^{tot}=N_{J/\psi}~+~R(\psi^{\prime})N_{\psi^{\prime}}~+~
~R(\chi_1)N_{\chi_1}~+~
R(\chi_2)N_{\chi_2}~,
\end{equation}
where $N_{J/\psi}$, $N_{\psi^{\prime}}$, $N_{\chi_1}$, $N_{\chi_2}$
are calculated according to Eq.(\ref{gce}) and $R(\psi^{\prime})\cong
0.54$, $R(\chi_1)\cong 0.27$, $R(\chi_2)\cong 0.14$ are the decay
branching
ratios of the excited charmonium states into $J/\psi$.

In the canonical ensemble (c.e.) formulation (i.e. when the requirement of
zero "charm charge"
of the HG is used in the exact form) the thermal charmonium
multiplicities are still given by Eq.(\ref{gce}) as charmonium states
have zero
charm charge.
The multiplicities (\ref{gce}) of open charm hadrons
will however be multiplied by an additional
`canonical suppression' factor (see e.g. \cite{Go1}).
This suppression factor is the same for all individual open 
single charm states.
Therefore, if
$N_O$ is the total g.c.e.
multiplicity of all open charm and anticharm mesons and (anti)baryons,
the c.e. value of the total open charm is equal to:
\begin{equation}\label{NOce}
N^{c.e.}_{O}~=~N_O~\frac{I_1(N_O)}{I_0(N_O)}~,
\end{equation}
where 
$I_0,I_1$ are the modified Bessel functions.
%$N_{H}$
%is the total
%thermal multiplicity of particles with hidden charm.
To find $N_O$ 
we use Eq.(\ref{gce}) for 
thermal multiplicities of the open charm hadrons in the g.c.e.
and take the
summation over all known particles
and resonances with 
open charm \cite{pdg}.
The canonical suppression factor $I_1(N_O)/I_0(N_O)$
in Eq.(\ref{NOce})
is due to the exact charm conservation.
Therefore,
the baryonic number, strangeness and electric 
charge of the HG system are treated according to the
g.c.e. but charm is considered in the c.e.
formulation where the exact charge conservation is imposed.
For $ N_O<<1$ one has
$I_1(N_O)/I_0(N_O)\cong N_O/2$ and,
therefore, the c.e. total open charm multiplicity is strongly suppressed
in comparison to the g.c.e. result.
At the SPS energies the c.e.
suppression effects
are important for the thermal open charm yield  even in
the most central Pb+Pb collisions. These  
suppression effects become
crucial when the number of participants $N_p$ decreases.
Note that for $N_O << 1$ the multiplicities of the open charm
hadrons are proportional to  $V^2$ in the c.e. HG formulation
(instead of $V$ in the g.c.e.).

The statistical coalescence model (SCM)
\cite{Br1,Go:00} assumes that the charmonium states
are formed at the hadronization stage.
This is similar to the thermal model of Ref.~ \cite{Ga1}.
The thermal model \cite{Ga1} predicts that the $J/\psi$ to
$\pi$ ratio is independent of $\sqrt{s}$ and $N_p$
at high collision energies. This is because both
$\langle J/\psi \rangle$ and $\langle \pi \rangle$
are proportional to the system volume and they both depend only
on the hadronization temperature, $T_H=170\pm 10$~MeV,
which is expected to be independent of $\sqrt{s}$ and $N_p$
at high collision energies. 
However, in the SCM the  charmonium states are 
produced via a coalescence of created
earlier $c$$\overline{c}$ quarks 
at the early
stage of A+A reaction by the hard parton collisions.
%The number of created $c\overline{c}$
%pairs, $N^{c\overline{c}}$, differs in general from the result
%expected in the  equilibrium HG.
One needs then an additional parameter
 $\gamma_c$  \cite{Br1} to
adjust the thermal HG results to the required number
of $N_{c\overline{c}}$.
This is analogous to the
introduction of the strangeness suppression factor $\gamma_s < 1$ \cite{Raf1} 
in the HG model, if the total strangeness observed is smaller than its 
thermal equilibrium
value. 
We find $\gamma_c>1$
so that the open charm hadron yield
is enhanced
by a factor $\gamma_c$ and charmonium  
yield by a factor $\gamma_c^2$ in comparison
with the equilibrium HG predictions.
The c.e. formulation of the SCM
is \cite{Go:00}:
\begin{equation}\label{Ncc1}
N_{c\overline{c}}~=~\frac{1}{2}~
\gamma_c~N_O~\frac{I_1(\gamma_c N_O)}{I_0(\gamma_cN_O)}~
+~\gamma_c^2~N_{H}~,
\end{equation}
where $N_{H}$
is the total HG
multiplicity of particles with hidden charm.
Note that the second term in the right-hand side of Eq.(\ref{Ncc1}) gives
only a tiny
correction to the first term, i.e.  most  of the created
$c\overline{c}$ pairs are transformed into the open charm hadrons.
If $N_{c\overline{c}} >> 1$ one finds from 
Eq.(\ref{Ncc1}) that $\gamma_c N_O>>1$, therefore,
$I_1(\gamma_cN_O)/I_0(\gamma_c N_O)\rightarrow 1$, i.e. the g.c.e. and
c.e. results for the open charm coincide.
In this case Eq.(\ref{Ncc1})
is simplified to \cite{Br1}:     
%$\begin{equation}\label{Ncc}
$N_{c\overline{c}}=\gamma_c~N_O/2+\gamma_c^2N_{H}$.
%\end{equation}
%For the lower RHIC energy $\sqrt{s}=56$~GeV  
%the value of
%$N^{dir}_{c\overline{c}}$
%is close to unity so that Eq.(\ref{Ncc1}) should be used
%instead of (\ref{Ncc}).
This happens for central  Au+Au collisions
at upper RHIC energy  $\sqrt{s}=200$~GeV.
For  lower RHIC energy $\sqrt{s}=56$~GeV
the value of
$N_{c\overline{c}}$
could be close to (or even smaller than) unity so that
the c.e. suppression effects for the open charm are
still important. Note that for the non-central A+A collisions
the c.e. suppression effects
become evidently stronger,
hence their consideration are necessary to study the
$N_p$ dependence of the charmonium production even
at upper RHIC energy.

Eq.(\ref{Ncc1}) will be used to find the charm enhancement
factor $\gamma_c$ and calculate then the $J/\psi$ multiplicity:
\begin{equation}\label{Npsi}
\langle J/\psi \rangle~= ~\gamma_c^2~N_{J/\psi}^{tot}~,
\end{equation}
where $N_{J/\psi}^{tot}$ is given by Eq.(\ref{psitot}).
Note that $T=170\div180$~MeV leads to the value of the thermal
ratio,
\begin{equation}\label{psi'}
\frac{\langle \psi^{\prime} \rangle } {\langle J/\psi \rangle}~=
~\left(\frac{m_{\psi^{\prime}}}{m_{J/\psi}}\right)^{3/2}
\exp\left(-~\frac {m_{\psi^{\prime}} - 
m_{J/\psi}}{T}\right) ~\cong~0.04 \div 0.05~,
\end{equation}
in agreement with data \cite{psi'} in Pb+Pb collisions at SPS for 
$N_p>100$. This fact was first noticed in Ref.~\cite{psi'1}.
Recent results \cite{mt} on transverse mass 
spectra of $J/\psi$ and $\psi^{\prime}$
mesons in central Pb+Pb collisions at 158 AGeV also support a hypothesis
of statistical production of charmonia at 
hadronization (see Ref.~\cite{mt1}).

The number of directly produced
$c\overline{c}$ pairs in the left-hand
side of Eq.(\ref{Ncc1}) should be estimated 
in the pQCD approach and 
used then as the input for the SCM.
The pQCD calculations for ${c\overline{c}}$ production
cross sections were
first done in Ref.\cite{comb}.  
For the cross section
$\sigma(pp\rightarrow
c\overline{c})$ of the charm production in p+p collisions we use 
the results presented in Ref.\cite{ruusk}.
This leads to the value of
$\sigma(pp\rightarrow
c\overline{c})~\cong 0.35$~mb at $\sqrt{s}=200$~GeV
and the $\sqrt{s}$-dependence of the cross section
for $\sqrt{s}=10\div200$~GeV is parameterized
as:
\begin{equation}\label{pert1}
\sigma(pp\rightarrow c\overline{c})~=~\sigma_0~\cdot 
\left(1- \frac{M_{0}}{\sqrt{s}}\right)^{\alpha}~
\left(\frac{\sqrt{s}}{M_{0}}\right)^{\beta}~,
\end{equation} 
with $\sigma_0 \cong 3.392 $~$\mu$b, $M_{0}\cong 2.984$~GeV, 
$\alpha \cong 8.185$ and $\beta \cong 1.132$.

The number of produced $c\overline{c}$
pairs in A+A collisions is proportional to
the number of primary N+N collisions, $N_{coll}^{AA}$,
which in turn is proportional to $N_p^{4/3}$ \cite{eskola}:
\begin{equation}\label{pert}
N_{c\overline{c}}~ = ~N_{coll}^{AA}(N_p)~
\frac{\sigma(pp\rightarrow c\overline{c})}{\sigma_{NN}^{inel}}~
\cong ~ C~ \sigma(pp\rightarrow c\overline{c})~N_p^{4/3}~,
\end{equation}
where $\sigma_{NN}^{inel}\cong 30$~mb is the inelastic N+N cross sections,
$C\cong 11$~barn$^{-1}$.

%To calculate the $J/\psi$ multiplicity (\ref{psitot})
%and the total open charm multiplicity $N^{c.e.}_{O}$
%(\ref{NOce}) in the HG model we need
%the hadronization parameters $V,T,\mu_B$ in A+A collisions.
%For the  RHIC energies the temperature parameter $T$ is expected to be
%similar to that for the SPS energies: $T=170\pm 10$~MeV.
%To fix the unknown
%volume parameter $V$ and baryonic chemical
%potential $\mu_B$ we use the parameterization of the 
%pion multiplicity per nucleon participant \cite{Ga:pi}:
%\begin{equation}\label{pionexp}
%\frac{\langle \pi \rangle}{N_p} ~\cong ~1.46~F~,
%\end{equation}
%where $F\equiv (\sqrt{s}-2m_N)^{3/4}/\sqrt{s}^{1/4}$.
%Eq.(\ref{pionexp}) is in agreement with both the SPS Pb+Pb data and
%preliminary RHIC data in Au+Au collisions at
%$\sqrt{s}=56, 130, 200$~GeV.
%The pion multiplicity $\langle \pi \rangle$ in Eq.~(\ref{pionexp}) 
%should be equaled
%to the total HG pion multiplicity $N_{\pi}^{tot}$
%calculated according to Eq.(\ref{dec}). 
%The HG parameters $V$ and $\mu_B$ are found then as the solution
%of the following coupled equations:
%\begin{eqnarray}
%\label{pi}
%\langle \pi \rangle ~ & = & N_{\pi}^{tot}(V,T,\mu_B)~\equiv~
%V~n_{\pi}^{tot}(T,\mu_B)~,\\
%~V~\left[~n_{\pi}(T)~+~ \sum_i B(i\rightarrow
%\pi) n_i(T,\mu_B)~\right]~,\\ 
% \label{Np}
%N_p~ & = & ~V~n_B(T,\mu_B)~,  
%\end{eqnarray}
%where $n_B$ is the HG baryonic density. In
%these calculations we fix $T\cong 170$~MeV.

The results of the SCM can be studied analytically in
the limiting cases of small and large numbers of 
$N_{c\overline{c}}$. For $N_{c\overline{c}}<<1$
one finds from Eq.~(\ref{Ncc1}) that 
$\gamma_c^2\approx
4N_{c\overline{c}}/N_O^2$ 
and then from Eq.~(\ref{psitot})
\begin{equation}\label{lim1}
R~
%\equiv~\frac{\langle J/\psi \rangle}{N_{c\overline{c}}}~
=~\frac{\gamma_c^2 N_{J/\psi}^{tot}}{N_{c\overline{c}}}~
\cong~  \frac{4 N_{J/\psi}^{tot}}{N_O^2}~\sim~\frac{1}{V}~\sim~
\frac{1}{\langle \pi \rangle} ~\sim~ (\sqrt{s})^{-1/2}~N_p^{-1}~.
\end{equation}
This behavior is similar to the standard picture
of the $J/\psi$ suppression: the ratio $R$ 
decreases with increasing both $\sqrt{s}$ and $N_p$.
 Therefore, the SCM predicts the
$J/\psi$ {\it suppression} at the SPS energy. This energy is still 
too ``low''
as $N_{c\overline{c}}<1$ even in the most central Pb+Pb
collisions at 158~AGeV. This behavior is however dramatically changed
at the RHIC energies \cite{Mc}. In most central Au+Au collisions ($N_p\cong 2A$)
at $\sqrt{s}=200$~GeV the expected number of $N_{c\overline{c}}$
is essentially larger than unity.  For $N_{c\overline{c}}>>1$
one finds from Eq.~(\ref{Ncc1}) that  
$\gamma_c\approx
2N_{c\overline{c}}/N_O$ 
and then from Eq.~(\ref{psitot})
\begin{equation}\label{lim2}
R~
%\equiv~\frac{\langle J/\psi \rangle}{N_{c\overline{c}}}~
=~\frac{\gamma_c^2 N_{J/\psi}^{tot}}{N_{c\overline{c}}}~
\cong~  \frac{2 N_{c\overline{c}}}{N_O}~\sim~\frac{N_{c\overline{c}}}{V}~\sim~
\frac{N_{c\overline{c}}}{\langle \pi \rangle} ~\sim~ (\sqrt{s})^{\beta 
-1/2}~N_p^{1/3}~,
\end{equation}
where $\beta\cong 1.132$. Eqs.~(\ref{lim1},\ref{lim2}) reveal a remarkable
prediction of the SCM: the $J/\psi$ to $N_{c\overline{c}}$ ratio 
decreases with both $\sqrt{s}$ and $N_p$ 
($J/\psi$ {\it suppression} (\ref{lim1})) when $N_{c\overline{c}}<<1$
and it increases with both $\sqrt{s}$ and $N_p$
($J/\psi$ {\it enhancement} (\ref{lim2})) when $N_{c\overline{c}}>>1$.
The measurements in Au+Au collisions at RHIC give a unique possibility 
to check simultaneously both these predictions:
at $\sqrt{s}=56$~GeV and $N_p=100$ the expected value of $N_{c\overline{c}}$
from Eq.~({\ref{pert}) is 
$N_{c\overline{c}}\cong 0.2<<1$, 
but at $\sqrt{s}=200$~GeV and $N_p\cong 2A$ one
expects from Eq.~({\ref{pert}) that 
$N_{c\overline{c}}\cong 10>>1$. Therefore, changing of $\sqrt{s}$ and $N_p$
in Au+Au at RHIC one could observe both the $J/\psi$ {\it suppression}
and {\it enhancement} behaviors.  These limiting behaviors are smoothly connected in
the intermediate region of  $\sqrt{s}$ and $N_p$ where $N_{c\overline{c}} \cong 1$.   
The results of the SCM for the $J/\psi$ to $N_{c\overline{c}}$ ratio  are presented in
Figs.~1 and 2.
Both the suppression (the dashed line in Fig.~3) and enhancement
(the solid lines in Figs.~2 and 3) behaviors are clearly seen.

\begin{figure}[ht]  
\begin{center}
\includegraphics[height=7.6cm]{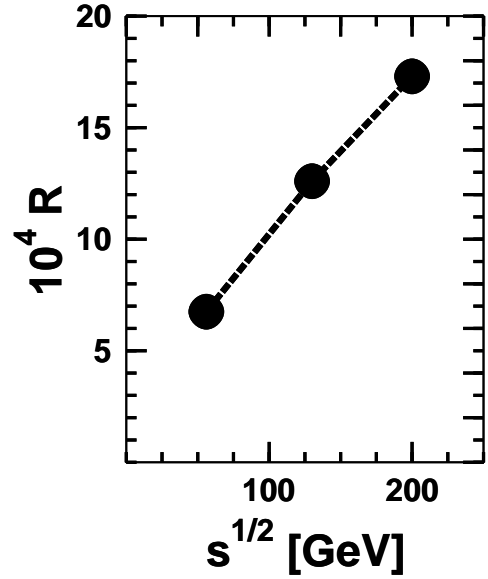}
\caption{The energy dependence of the $J/\psi$ to
$N_{c\overline{c}}$
ratio in central ($N_p\cong 2A$) Au+Au collisions.
Points are the predictions of the SCM
for the RHIC energies: $\sqrt{s}=56, 130, 200$~GeV.
The ratio $R$ increases by a factor of
about 3 ($J/\psi$ {\it enhancement}) in the region of RHIC energies
$\sqrt{s}=56\div 200$~GeV.}
\end{center}
\end{figure}

\begin{figure}[ht]
\begin{center}
\includegraphics[height=7.6cm]{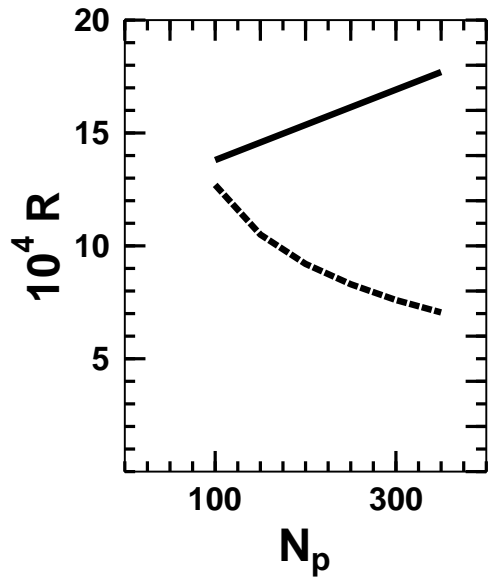}
\caption{The $N_p$-dependence of the $J/\psi$ to
$N_{c\overline{c}}$
ratio. The lines are the predictions of the SCM. The dashed line corresponds
to  $\sqrt{s}=56$~GeV and shows the $J/\psi$ {\it suppression} behavior.
The solid line corresponds   
to  $\sqrt{s}=200$~GeV and shows the $J/\psi$ {\it enhancement}.
}
\end{center}
\end{figure}

\section{ Conclusions}

\begin{itemize}

\item
The deconfinement phase transition is expected to occur at the
collision energies between AGS and SPS
where the strangeness to entropy (pion) ratio reveals
a non-monotonic (or kinky) dependence
on the collision energy \cite{early} (see Fig.~1).

\item Statistical hadronization of the
QGP is probably an important
source of $J/\psi$ production \cite{Ga1}.
This fact would open a new look at $J/\psi$ `suppression'
signal of the QGP.
The statistical coalescence model \cite{Br1,Go:00} 
of the $J/\psi$ production
predicts that the $J/\psi$ {\it suppression}
in peripheral Au+Au collisions at lower RHIC energy
should be changed into the $J/\psi$
{\it enhancement} \cite{Mc} in central 
Au+Au collisions at the upper RHIC energy (see Figs.~2 and 3).

\end{itemize}

%{\bf  Acknowledgments.}  
\ack
I am thankful to K.A.~Bugaev, M. Ga\'zdzicki,
W. Greiner, A.P.~Kostyuk, L.~McLerran and H.~St\"ocker for fruitful
collaboration. I am also 
thankful to F. Becattini, P. Braun-Munzinger, J.~Cleymans, 
L. Gerland, D.~Kharzeev, I.N.~Mishustin, G.C.~Nayak,  K.~Redlich and J.~Stachel for
comments and
discussions.
The research described in this publication was made possible in part by
Award \# UP1-2119 of the U.S. Civilian Research and Development
Foundation for the Independent States of the Former Soviet Union
(CRDF).

\section*{References}
%\begin{harvard}

%\end{harvard}
%\endrefs

\end{document}